# Measuring the Impact of Spammers on E-Mail and Twitter Networks

Fronzetti Colladon, A., & Gloor, P. A.

# Measuring the Impact of Spammers on E-Mail and Twitter Networks

Fronzetti Colladon, A., & Gloor, P. A

**Abstract**

This paper investigates the research question if senders of large amounts of irrelevant or unsolicited information – commonly called "spammers" - distort the network structure of social networks. Two large social networks are analyzed, the first extracted from the Twitter discourse about a big telecommunication company in Italy, and the second obtained from three years of email communication of 200 managers working for a large multinational company. This work compares network robustness and the stability of centrality and interaction metrics, as well as the use of language, after removing spammers and the most and least connected nodes. The results show that spammers do not significantly alter the structure of the information-carrying network, for most of the social indicators. The authors additionally investigate the correlation between e-mail subject line and content by tracking language sentiment, emotionality, and complexity, addressing the cases where collecting email bodies is not permitted for privacy reasons. The findings extend the research about robustness and stability of social networks metrics, after the application of graph simplification strategies. The results have practical implication for network analysts and for those company managers who rely on network analytics (applied to company emails and social media data) to support their decision-making processes.

**Keywords:** data preprocessing, robustness, social network, spammers, stability, email, Twitter.

## 1. Introduction

In today's era of constant availability of all types of online social media, it has become too easy to pull the trigger on social media. Rapidly growing social platforms, such as Facebook and Twitter, have been widely infiltrated by spam messages (Wang, 2010), with associated cleanup- and filtering costs arising, for example, from the proliferation of fake news (Marchi, 2012). Inside the company, emails are still a major communication channel, as they provide a fast, user friendly and cost-effective communication. The management of emails, and the problem of email overload, remain a major challenge for organizations (Sumecki, Chipulu, & Ojiako, 2011). Today's companies well understand that interrupting employees through too many emails has a cost: it can lead to stress and frustration (Mark, Gudith, & Klocke, 2008), as well as to decreasing work efficiency (Jackson, Dawson, & Wilson, 2001). Indeed, a large amount of the emails generated within organizations is not business critical (Sumecki et al., 2011). Employees are often subjected to a constant firestorm of e-mails from the CEO and senior management, asking for their commitment to the companies' objectives, or sharing recent achievements and future goals. Human resources managers send messages about changing rules and regulations, and well-meaning volunteers solicit donations for United Way and other charities. In this paper, the authors investigate different methods to reduce this cacophony of online chatter to be able to better analyze the underlying social network structure, dynamics, and content. Reducing the disturbance in email communication can help social network analysts, and support better interactions among employees, or between a company and its clients.

E-mail and online social networks provide an unprecedented view into the inner workings of organizations by providing access to their internal and external communication behavior (Gloor, 2005, 2017). Using dynamic social network analysis, researchers can gain deep insights into the

functioning of these networks. Analyzing the email communication patterns within large multinational companies, for example, proved its value in predicting managerial turnover (Gloor, Fronzetti Colladon, Grippa, & Giacomelli, 2017) or behaviors that can improve customer satisfaction (Gloor, Fronzetti Colladon, Giacomelli, Saran, & Grippa, 2017). However, depending on the algorithms used, analyzing large datasets can be very time consuming or require costly hardware solutions (Gandomi & Haider, 2015; Kaisler, Armour, Espinosa, & Money, 2013; Yaqoob et al., 2016). In order to reduce these problems, data cleaning and simplification techniques can be implemented, to reduce the sample size and consequently the algorithmic computation times, and to study a "cleaner" signal (Rahm & Do, 2000; Stringhini, Kruegel, & Vigna, 2010). In the process of data cleaning, the analyst has the duty to correct or eliminate inaccurate or incomplete records, as well as irrelevant observations, which might represent a disturbance of the general insights coming from the analysis. "Spammers" inside and outside of the organization potentially influence the social network structure, and thus degrade the value of predictive analytics (Bonchi, Castillo, Gionis, & Jaimes, 2011). Therefore, they represent a cluster of social actors which the analyst might want to remove before carrying out a network analysis – with the double advantage of 'cleaning' the insights provided by social interaction metrics and reducing the computational complexity of the analysis. This intent is supported by a large body of literature which already proposed methods and tools to effectively recognize spammers (Zheng, Zeng, Chen, Yu, & Rong, 2015) and filter spam messages, both in email (Bhowmick & Hazarika, 2018) and in social media communication (Xu, Sun, & Javaid, 2016). Nonetheless, few studies analyzed the impact of node and link removal strategies on the structure of the communication network (e.g., Costenbader & Valente, 2003; Fronzetti Colladon & Vagaggini, 2017). In contexts where spam detection is more difficult, or too time consuming, researchers might ask themselves if leaving spammers in the network will significantly affect the results of their investigation. Alternatively, when spammers can be correctly identified, researchers might wonder if removing them could significantly alter the social network structure and consequently the results of the analysis. Therefore, this work focuses on the research question: "do spammers influence the social network structure?".

Removing spammers can be considered a graph simplification strategy; however, spam identification is not always easy, especially in contexts where message content is not available to the analyst, for example for privacy reasons (e.g., Gloor, Fronzetti Colladon, Giacomelli, et al., 2017). Indeed, many filtering algorithms are based on content analysis (Bhowmick & Hazarika, 2018); when this is not possible, other network-based strategies have to be employed (Zheng et al., 2015), such as the identification of overly connected or very peripheral nodes. Even if these strategies are less effective, they are sometimes considered as a viable alternative (Fronzetti Colladon & Vagaggini, 2017). Accordingly, this work explores the impact of other graph simplification techniques and compares them with the direct removal of spammers.

The authors conducted their experiments considering two real world networks: the first extracted from Twitter and the second obtained fetching three years of email communication within a large company operating in the utilities sector. The authors refer to metrics and methods from the field of Social Network Analysis (Borgatti, Everett, & Johnson, 2013), combining them with an investigation of the network dynamics (such as the response times) and with variables related to the use of language (sentiment, emotionality and complexity). Combining these three level of analysis offers deeper insights than just focusing on a single dimension of the communication process (Gloor, Fronzetti Colladon, Giacomelli, et al., 2017). More specifically, the authors start with the identification of the "spammers" in their case study networks: those social actors who send mostly undesired and irrelevant messages (see the identification details in Section 3). Subsequently, they test the stability of different network metrics while removing the spammers and the least connected nodes, which are located at the network periphery. The authors also investigate the difference of the impacts produced by the removal of actors suspected to be spammers just because they are overly connected in the graph. The aim is to empirically explore the effects produced by various data cleaning strategies applied to real communication networks and to provide suggestions to the data scientists on tests to carry out before choosing the most appropriate strategy. From the hundreds of available network metrics and data cleaning strategies, the authors made a selection of some which are well-known and had been identified to be particularly useful in previous studies (Fronzetti Colladon & Vagaggini, 2017; Wasserman & Faust, 1994).

This work extends the research about social network and text mining metrics, proposing a methodology to assess their robustness and stability in the presence of spammers. These metrics often represent 'honest signals' of collaboration (Pentland, 2008), whose stability has not yet been extensively tested when applying graph simplification techniques. The authors' findings are similarly useful for researchers and business analysts involved in the study of online interaction and digital communication. Using the approach presented in this paper, analysts can include potential effects produced by node removal into their analysis. 'Simplifying the network graph, before carrying a social network analysis, is indeed often useful to reduce the computational complexity of some algorithms and to clean the signal from possible disturbances' (Fronzetti Colladon & Vagaggini, 2017, p. 1288). In addition, understanding the impact of spammers can support managers who want to improve communication dynamics within their companies, or who are concerned about problems of communication overload.

## 2. Network Robustness and Stability

Network *resilience* can be considered as the ability of a system to return to its normal state after disturbance, based on early detection and fast recovery, while network *robustness* is the ability of a network to retain its system structure even when exposed to perturbations, i.e. while removing a subset of nodes and links (Holmgren, 2007). This paper is focused on the concept of network robustness and tests the effects of node deletion for data cleaning purposes. More specifically, the authors consider static changes in network structure, as well as the stability of a set of important node-level metrics.

Newman (2010) presented different criteria for how to select ties and actors to remove from a social network, for instance degree and betweenness centrality, with the aim of testing system robustness. In general, it is possible to distinguish between random failures, which can affect any node, and targeted strategies, where nodes to remove are chosen according to their importance in the network, usually with the purpose of reducing network connectivity. Targeted attacks require at least some prior knowledge of the network structure. In addition, interventions can be simultaneous or sequential, taking into account the system's reaction before deciding on the next node to target (Holme, Kim, Yoon, & Han, 2002). The authors' experiment is focused on simultaneous selections, as the graph simplification that the analyst carries out will happen all at once.

Past research has investigated the properties of robustness and resilience of complex networks often in relation to events that could harm the network connectivity, classified as random failures or targeted attacks on specific nodes and ties. While the research question of investigating changes in structural network properties through removal of spammers is somewhat different from the discussion on network resilience under malicious attack, it nevertheless informs our research. The understanding of vulnerabilities and of network stability can be extremely useful both for the design of robust systems (Estrada, 2006; Sohal, Sandhu, Sood, & Chang, 2018) and to gain insights which can help anticipate the effects of graph simplification choices (even if such choices are not directly classifiable as network attacks). Albert and colleagues (2000) studied the effects of node deletion on a sample made of webpages with cross links, finding that random failures have a small impact on increasing the average distance between nodes, whereas attacks directed to hubs can have a disruptive effect. This result also depends on the network structure of the World Wide Web, which has a degree distribution that can be approximated with a power-law (Broder et al., 2000). In the same work, authors proved that the effects of both attack strategies are small if the networks are not scale-free, but more homogeneously wired (Albert et al., 2000). Similar conclusions are drawn by Callaway et al. (2000) who tested network stability by means of percolation models and by Xiao et al. (2008) who showed that attacks in communication networks are more dangerous when the attacker has prior knowledge of the network structure. Many real world communication networks are indeed scale-free, with a small number of highly connected nodes and a large number of nodes with a much lower degree (Barabasi, 2002). Comparable results were also found by Broder et al. (2000) who proved that the World Wide Web is more robust than expected because to destroy its connectivity an attacker would have to remove all the vertices with a degree greater than five. Another important element of the network topology, which is linked to its robustness and to the possible diffusion dynamics, is the presence of a giant component, a larger connected subgraph that contains a significant fraction of nodes. The reduction in size of the giant component is a good proxy of the damage produced after an attack: if the giant component still exists, with a relatively small reduction in size, then the attack can be considered less harmful (Chen & Cheng, 2015; Holme et al., 2002;

Iyer, Killingback, Sundaram, & Wang, 2013). Besides looking at the giant component and the average distance between nodes, other measures can be used to test the robustness at the network level, such as centrality measures or clustering coefficients (Cohen & Havlin, 2010). Shargel, Sayama, Epstein and Bar-Yam (2003) studied the variation of the network diameter (the maximum length of the shortest paths connecting reachable pairs) after attacks on scale-free and exponential random graphs (Lusher, Koskinen, & Robins, 2013). Assortativity (degree-degree correlation of adjacent nodes) has been related to the concept of robustness, maintaining that a network core can better resist attacks when the assortative coefficient is higher (Rubinov & Sporns, 2010). Iyer et al. (2013) proved that network robustness is both dependent on the clustering coefficient and the level of assortativity. Similarly, Trajanovski et al. (2013) showed that increasing assortativity is useful against targeted attacks, but less relevant when facing random failures. Other scholars addressed the problem of finding reasonable defense strategies against targeted attacks, based, for instance, on the addition of links to average degree nodes (Yehezkel & Cohen, 2012) or on sequential defensive interventions (Chen & Cheng, 2015).

At the node level, Fronzetti Colladon and Vagaggini (2017) presented research which tests the effects of removing moderators from intranet online forums. Costenbader and Valente (2003) examined the stability of common centrality measures, calculated for each node of the original network as well as for random samples extracted from the same network. This paper presents a similar experiment, except that node selection is not random – resembling graph simplification strategies, based on the removal of spammers – and that two additional dimensions of the communication process are considered, namely the use of language and interaction dynamics.

To the extent of the authors' knowledge, few extant studies address issues related to graph simplification for large social networks with the purpose of removing unwanted actors, such as spammers or email accounts of mailing lists (e.g., Zilli, Grippa, Gloor, & Laubacher, 2006). Therefore, this study is meant to offer a contribution to this field.

3. **Data Collection and Analysis**

The case study started with the collection of data from two real world networks involving two large multinational companies. Firstly, the authors considered an email network representative of the communication patterns of about 200 managers working for a large company operating mostly in the US. More than 260,000 email messages were collected over a period of three years, enabling the authors to generate a directed graph with more than 27,000 nodes (email accounts). In this network, there is a link between two email accounts if they exchanged at least one email message during the study period. Secondly, the authors extracted a Twitter network made of about 17,000 nodes (Twitter accounts) and more than 80,000 links. This network was collected fetching tweets with content relevant to a large multinational company operating in the telecommunication industry, over a period of two months. Two Twitter accounts were connected by a link if they were retweeting or mentioning each other, or if they were answering to each other's tweets (see also Figure 1).

Twitter is a microblogging platform with more than 330 million active users[1], which has been included in this study for two main reasons: its API system, which provides free access to researchers, and the huge attention it receives from scholars and practitioners. Google Scholar lists in its index more than 1,990,000 studies that mentioned the word 'Twitter' in the last ten years.

---

[1] https://www.statista.com/statistics/282087/number-of-monthly-active-twitter-users/, accessed July 2, 2018.

Both networks present a degree distribution which is approximately scale-free (Barabasi, 2002), i.e. they are made of many nodes with few connections and a few hubs with a much larger number of links, exchanging emails or tweets with a large number of peers.

Some previous studies addressed the problem of network robustness adopting a simulative approach (e.g., Booker, 2012), i.e. testing the effect of nodes or tie removal on simulated random graphs (with varying degree distributions and structural properties). Simulations are useful as they allow users to test different effects of removal strategies on different graph structures. However, when performing targeted selections on simulated graphs, nodes to remove are usually selected for their higher centrality. The authors' choice was to analyze fewer real world networks, as one of the objectives of this paper is to specifically test the effect of the removal of spammers – whose selection can be very accurate when analyzing real email or Twitter accounts. In this work, "spammer" nodes are those social actors who are typically spamming the network with undesired and mostly unread messages, which might be advertising messages, real spam or phishing attempts, bots on Twitter, or just mailing lists internal to a business context. Consistent with previous studies (Stringhini et al., 2010), the authors identified spammers as the nodes which met the minimum number of conditions reported in Table 1.

| **Email Network** | **Twitter Network** |
|---|---|
| **(meets at least two conditions)** | **(meets at least three conditions)** |
| A) Sends a large number of messages to other email accounts (high volume). | A) Send a very high number of tweets, sometimes all through the 24 hours. |
| B) Receives zero or very few messages by other email accounts (not identified as spammers). | B) Is never, or very rarely, mentioned in the tweets of other accounts (not identified as spammers). |
| C) Messages content is manually classified as spam based on its content (e.g., advertising messages, phishing attempts). | C) Message content is manually classified as spam since clearly an advertising or phishing attempt, or because messages always contain a high number of links to external websites (Stringhini et al., 2010). |
| | D) Is following a much larger number of accounts than the number of its followers. |

**Table 1.** Criteria for the identification of spammers.

In such a context, it may happen that an email account has a high number of contacts or emails received, just because it receives a lot of spam or because it is a member of several mailing lists. Manual detection of spammers, which was carried out in the experiment, can be rather time consuming. For large networks, supervised machine learning algorithms can automatically identify nodes that behave as spammers by the content of their messages or by their network positions (Ramachandran & Feamster, 2006; Stringhini et al., 2010).

In addition, before carrying out a full network analysis, it might be sometimes useful to remove the "noise" generated by the nodes located at the network periphery – including isolates, and those nodes with just a single link or with a very low degree. Depending on the specific goal and on the context, different strategies will be used: the objective of this research is not to comment on the possible data cleaning strategies, but to explore the effects produced by the removal of selected nodes both on individual metrics (ego-network) and at the whole network level. Consistently with this premise, different selection methods were tested, removing:

a) the spammers to clean the network from unsolicited and mostly unwanted links;
b) the nodes in the top first, fifth and tenth percentile of the network degree distribution, mostly for comparative purposes and to see the effects produced by a removal strategy based on the selection of nodes with a high degree centrality (which can sometimes be useful if spammers can only be recognized as the overly connected nodes);
c) the nodes at the bottom of the network degree distribution, i.e. the isolates and those nodes connected by a single link, for their less significant role in large networks;
d) the combination of the nodes in the top and bottom selections, as a strategy to apply when the identification of spammers is not possible;
e) the combination of bottom nodes and spammers, to get rid both of unimportant nodes and manually identified spammers.

### 3.1. Variables description

While removing nodes from the two networks, the authors considered the effects produced both on whole network metrics – such as the diameter, or the clustering coefficient – and on individual metrics for the remaining actors – such as their degree or betweenness centrality. In addition, the impact on the sentiment, the emotionality and the complexity of the language used was explored (Brönnimann, 2014).

Networks were represented as directed graphs with *n* nodes (Twitter or email accounts) – denoted as G = {$g_1$, $g_2$, $g_3$ … $g_n$} – and of *m* directed arcs (emails; retweets, mentions, answers to tweets) linking these nodes. These graphs can be expressed by an adjacency matrix X of n rows and n columns, where the element $x_{ij}$ positioned at the row *i* and column *j* is bigger than 0 if, and only if, there is a tie originating from the node $g_i$ and terminating at the node $g_j$.

### 3.1.1. Whole network metrics

*Average Distance Among Reachable Pairs (ADARP).* This measure considers the average length of the shortest paths linking every possible pair of nodes in the graph – where nodes can reach one another, i.e. when the shortest path has a length lower than infinite (Doreian, 1974).

*Diameter.* The network diameter represents the length of the shortest path linking the two most distant nodes in the network (considering reachable pairs).

*Clustering Coefficient.* Is a measure of the tendency of nodes to cluster together, forming densely connected groups, with a lower number of cross-group connections. This coefficient is measured as suggested by Watts and Strogatz (1998), counting the number of fully connected nodes triplet and dividing it by the number of triplets where one link is missing.

*Average Degree.* Is the average of the number of arcs incident in each node of the network.

### 3.1.2. Node level metrics

*Degree Centrality.* This measure counts the number of arcs connecting a node to its neighbors.

*Closeness Centrality.* This variable is calculated as the inverse of the distance of a node from all others in a network, considering the shortest paths that connects each pair of nodes. It is often used as a proxy for the speed by which a social actor can reach the others (Wasserman & Faust, 1994).

*Betweenness Centrality.* This measure counts the number of times a node is in-between the shortest paths that connect every other pair of nodes (Wasserman & Faust, 1994).

*Betweenness Oscillations.* This variable counts the oscillations in betweenness centrality for the node $g_i$, within a specific time interval. If the node is keeping its position with respect to the other nodes, then the betwenness oscillations are equal to zero. On the other hand, if the node is changing its value of betweenness centrality reaching local minima or maxima, then the betwenness oscillations will be bigger than zero. This measure, also known as *rotating leadership*, has been introduced by Kidane and Gloor (2007) and has been previously used both at the individual and at the team level, to predict group creativity (Kidane & Gloor, 2007), innovative potential of startups (Allen, Gloor, Fronzetti Colladon, Woerner, & Raz, 2016) and business success in collaborative innovation (Davis & Eisenhardt, 2011).

*Activity.* In the email network it counts the number of messages sent by a specific actor. In the Twitter network it corresponds to the number of tweets posted by a user.

*Contribution Index.* This measure is meant to express the unbalance in messages sent when compared to messages received by a social actor. The contribution of the node $g_i$ is defined by the formula:

$$\text{CI}(g_i) = \frac{\text{Messages Sent} - \text{Messages Received}}{\text{Messages Sent} + \text{Messages Received}}$$

It varies between [1,-1], where a contribution index of 1 identifies people who only send messages (or tweets) without receiving an answer (Gloor, Laubacher, Dynes, & Zhao, 2003). Such a high contribution index value sometimes identifies spammers. Even if spammers selection was carried out manually in this study, the authors noticed that spammers often are users who send a significant number of messages, together with a contribution index bigger or equal to 0.8.

*Average Response Time (ART).* This measure evaluates the average time it takes a social actor to respond to the email messages he/she receives, or to tweets which are directed at him (*Ego ART*). A second measure – *Alter ART* – measures the time it takes for others to respond to a user or to tweets where a user is mentioned (Gloor, Almozlino, Inbar, Lo, & Provost, 2014).

*Nudges.* This variable counts, on average, the number of pings (nudges) required before a social actor answers to an email or to a tweet directed to him (Gloor et al., 2014). Nudges can be subsequent emails sent to an employee who has not yet responded, or new tweets which keep mentioning a Twitter user before receiving an answer. Similarly to ART, this measure can be relative to the nudges sent by the node $g_i$ (*Ego Nudges*), or to the nudges sent from all other nodes to the node $g_i$ (*Alter Nudges*).

### 3.1.3. Semantic variables

*Sentiment* measures positivity of the language used in the emails or in the tweets sent by a user. This metric varies from [0,1], where 1 represents a positive sentiment, values around 0.5 express a neutral sentiment and lower values represent a negative sentiment. This variable was calculated using a multi-lingual classifier based on a machine learning method trained on large datasets extracted from Twitter (Brönnimann, 2014), by means of the software Condor[2].

*Emotionality.* This variable measures the level of emotionality of the language used, calculated as the deviation from neutral sentiment (Brönnimann, 2014). If the level of emotionality is high, there is a more vivid debate, with possibly the alternation of messages with positive and negative sentiment.

*Complexity* represents the average complexity of the vocabulary used and is calculated as the likelihood distribution of words within a text – i.e., the probability that each word of a dictionary appears in the text (Brönnimann, 2014). A word is considered as more complex when it appears rarely in the context analyzed, and not when just rare in general. This way, even very specific and rare words are considered as common if frequently used in a discourse.

To calculate the presented measures, and to fetch the network data from emails and Twitter, the authors used the social network analysis software Condor and the software Pajek (De Nooy, Mrvar, & Batagelj, 2012). Many other metrics, both from Social Network and Semantic Analysis could have been included in the study. Selection was based on two criteria: firstly the authors wanted to include metrics which are commonly used by network and semantic analysts, such as the sentiment of the language used, or degree and betweenness centrality (e.g., Freeman, 1979); secondly, they aimed to test the effects of nodes' removal on less-known metrics – such as betweenness oscillations – which however revealed high explanatory power in previous studies (Allen et al., 2016; Kidane & Gloor, 2007).

---

[2] http://www.galaxyadvisors.com/products/

## 4. Results

The node removal strategies presented in Section 3 were applied to both the Twitter and the email network. Table 2 shows the effects of node removal on the whole network metrics.

| **Twitter Network** | **ADARP** | **CC** | **AD** | **D** |
| --- | --- | --- | --- | --- |
| Full Network | 4.33 | 0.40 | 8.3 | 16 |
| Removed Spammers | 4.77 | 0.39 | 7.21 | 16 |
| Removed Bottom Nodes | 4.25 | 0.42 | 8.82 | 16 |
| Removed Top 1st Percentile | 4.54 | 0.34 | 5.48 | 14 |
| Removed Top 5th Percentile | 2.42 | 0.33 | 4.09 | 7 |
| Removed Top 10th Percentile | 2.36 | 0.32 | 3.45 | 6 |
| Removed Top 1st and Bottom Nodes | 4.48 | 0.36 | 5.72 | 14 |
| Removed Spammers and Bottom Nodes | 4.70 | 0.41 | 3.80 | 16 |
| **Email Network** | **ADARP** | **CC** | **AD** | **D** |
| Full Network | 4.20 | 0.24 | 18.29 | 14 |
| Removed Spammers | 4.72 | 0.16 | 10.14 | 15 |
| Removed Bottom Nodes | 3.91 | 0.29 | 39.59 | 13 |
| Removed Top 1st Percentile | 6.36 | 0.10 | 6.22 | 24 |
| Removed Top 5th Percentile | 9.07 | 0.06 | 2.76 | 39 |
| Removed Top 10th Percentile | 1.36 | 0.05 | 1.92 | 7 |
| Removed Top 1st and Bottom Nodes | 6.04 | 0.12 | 10.6 | 20 |

| Removed Spammers and Bottom Nodes | 4.47 | 0.20 | 7.5 | 13 |

*Notes.* ADARP = Average distance among reachable pairs; CC = Clustering coefficient; D = Network diameter; AD = Average degree.

**Table 2.** Impact of node removal on whole network metrics.

Figure 1 shows the network pictures, after having applied each removal strategy separately.

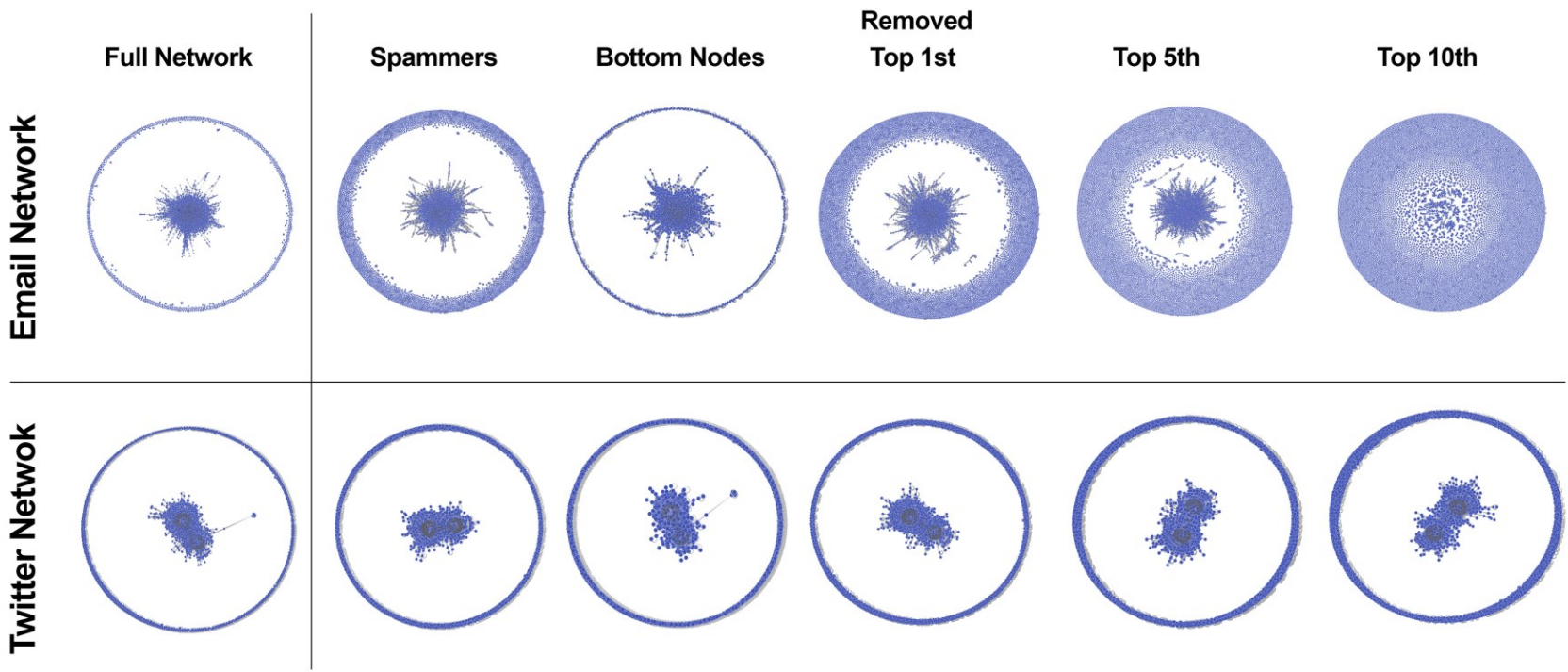

**Figure 1.** Network pictures after node removal.

Both networks are characterized by having a large periphery, with loosely connected nodes – i.e. isolates or nodes with just a single connection. However, the removal of "bottom" nodes (the "leaves" in the graph) does not seem to greatly affect the whole network metrics, apart from increasing the average degree. Similarly, removing the spammer nodes from the Twitter network did not have a big impact – the diameter remained unchanged and the other metrics had a reasonably small variation. By contrast, there were more significant variations in the email network: removing the spammers decreased the clustering coefficient and the average degree and slightly increased the network diameter. However, these variations were less significant than when removing the top connected nodes. In both networks, when removing the top nodes (especially when considering the nodes included in the top $5^{th}$ or $10^{th}$ percentiles), there were significant variations in all the metrics, i.e. a larger change in the connection patterns linking the social actors. This is also confirmed by the network pictures, which show that when removing the top connected nodes, the giant component is more significantly reduced – with a larger number of nodes relocated at the loosely connected network periphery, which surrounds the network core. When the network diameter significantly dropped under the value of 8, the size of the biggest network component was strongly reduced, as well as was the number of reachable pairs (having many small clusters or isolated nodes at the network periphery). Therefore, while it still seems reasonable for the analyst to remove spammers and peripheral nodes, it seems very dangerous to remove overly connected nodes since this can strongly impact network structure and metrics.

Tables 3a and 3b show the effects of node removal at the individual level. In order to make the tables easier to read, the authors chose to present, in each row, the correlation of the row variable with the full network.

| Twitter Network | Removed Nodes | | | | | | |
|---|---|---|---|---|---|---|---|
| | Spammers | Bottom Nodes | Top 1st Percentile | Top 5th Percentile | Top 10th Percentile | Spammer and Bottom Nodes | Top 1st Percentile and Bottom Nodes |
| **Alter ART** | .985* | 1.000* | .987* | .993* | .997* | .985* | .987* |
| **Ego ART** | .967* | 1.000* | .945* | .964* | .983* | .967* | .945* |
| **Alter Nudges** | .998* | 1.000* | .977* | .957* | .991* | .997* | .976* |
| **Ego Nudges** | .995* | 1.000* | .902* | .975* | .941* | .995* | .902* |
| **Activity** | .999* | 1.000* | .991* | .997* | .998* | .998* | .991* |
| **Contribution Index** | .961* | .997* | .943* | .920* | .904* | .950* | .938* |
| **Betwenness Centrality** | 1.000* | 1.000* | 1.000* | .999* | .999* | .999* | .999* |
| **Betwenness Centrality Oscillations** | .989* | .993* | .941* | .920* | .907* | .980* | .930* |
| **Closeness Centrality** | .871* | 1.000* | .799* | .733* | .691* | .909* | .877* |
| **Degree Centrality** | 1.000* | 1.000* | 1.000* | 1.000* | 1.000* | 1.000* | 1.000* |
| **Sentiment** | .984* | .987* | .974* | .959* | .948* | .982* | .973* |
| **Emotionality** | .964* | .982* | .949* | .928* | .909* | .961* | .946* |
| **Complexity** | .977* | .983* | .961* | .941* | .920* | .974* | .961* |
| * $p < .01$. | | | | | | | |

**Table 3a.** Correlation coefficients with original node level metrics for the Twitter network.

| Email Network | Removed Nodes | | | | | | |
|---|---|---|---|---|---|---|---|
| | Spammers | Bottom Nodes | Top 1st Percentile | Top 5th Percentile | Top 10th Percentile | Spammer and Bottom Nodes | Top 1st Percentile and Bottom Nodes |
| **Alter ART** | .947* | 1.000* | .906* | .917* | .968* | .945* | .905* |
| **Ego ART** | .938* | 1.000* | .881* | .905* | .974* | .936* | .879* |
| **Alter Nudges** | .948* | 1.000* | .849* | .905* | .949* | .948* | .848* |
| **Ego Nudges** | .961* | 1.000* | .803* | .910* | .972* | .961* | .802* |
| **Activity** | .987* | 1.000* | .888* | .728* | .625* | .986* | .876* |
| Contribution Index | .871* | .980* | .753* | .634* | .576* | .835* | .746* |
| Betwenness Centrality | .987* | .981* | .767* | .373* | .075* | .962* | .698* |
| Betwenness Centrality Oscillations | .950* | .997* | .870* | .500* | .309* | .945* | .857* |
| Closeness Centrality | .484* | .975* | .319* | .186* | .054* | .477* | .328* |
| Degree Centrality | .994* | .986* | .959* | .770* | .600* | .981* | .955* |
| Sentiment | .909* | .970* | .734* | .614* | .552* | .877* | .740* |
| Emotionality | .891* | .970* | .736* | .612* | .534* | .805* | .655* |
| Complexity | .901* | .970* | .742* | .619* | .539* | .829* | .681* |
| * p < .01. | | | | | | | |

**Table 3b.** Correlation coefficients with original node level metrics for the Email network.

Looking at the node-level metrics is often important as the structural positions and roles of social actors can be used by many studies and predictions (e.g., Borgatti et al., 2013; Cross & Prusak, 2002). Interaction metrics (Ego ART and Nudges, Alter ART and Nudges) are quite stable in both networks for all removal strategies used, being highly correlated with the original network values. Activity and Contribution index are relatively stable in the Twitter network, whereas these values are significantly affected when removing the top nodes from the email network – this is probably due to the fact that this network has a more connected core with a loosely connected periphery and this core is held together by the overly connected nodes. This could also be the reason why degree and betweenness centrality and betwenness oscillations show a similar stability in the Twitter network, but are unstable when removing top nodes from the email graph. Closeness centrality, on the other hand, seems to be significantly affected by every removal choice, except when removing the bottom nodes. The robustness of semantic variables is shown in the table for the Twitter network, while we find progressive change when removing the top nodes in the email network. Identifying the spammer nodes has less impact than just removing the top connected nodes, as all the metrics – both at the ego-level and the whole network level – are less affected by this. Consistently, a good strategy seems to be to remove the spammers and the bottom nodes from the analyzed networks, with the dual benefit of "cleaning" the analysis from disturbance and of simplifying the graph, without significantly harming the robustness of the network metrics (excluding closeness in the email experiment).

As a last step, the authors compared the sentiment, emotionality and complexity of the emails' body with their subject lines. Indeed, it is often the case that the analyst cannot access email bodies during data collection (Gloor, Fronzetti Colladon, Giacomelli, et al., 2017; Gloor, Fronzetti Colladon, Grippa, et al., 2017) for privacy reasons. Email bodies often contain more sensible information that a company or an individual is willing to share; privacy is preserved better when only sharing subject lines. Table 4 shows the results obtained for the email network, correlating the semantic variables both at the author level (the average sentiment/emotionality/complexity of subject lines and email bodies for each specific actor) and at the email level (the correlation of the sentiment/emotionality/complexity of the subject line and of the body of each email).

| Author Level (N=27,597) | | 1 | 2 | 3 | 4 | 5 | 6 |
|---|---|---|---|---|---|---|---|
| 1 | **Author Sentiment (Emails Content)** | 1.00 | | | | | |
| 2 | **Author Sentiment (Subject Lines)** | .25** | 1.00 | | | | |
| 3 | **Author Complexity (Emails Content)** | .33** | .22** | 1.00 | | | |
| 4 | **Author Complexity (Subject Lines)** | .25** | .26** | .74** | 1.00 | | |
| 5 | **Author Emotionality (Emails Content)** | -.26** | -.22** | -.90** | -.73** | 1.00 | |
| 6 | **Author Emotionality (Subject Lines)** | -.21** | -.16** | -.65** | -.81** | .66** | 1.00 |

| Email Level (N= 261,559) | 1 | 2 | 3 | 4 | 5 | 6 |
|---|---|---|---|---|---|---|
| 1 Email Content Sentiment | 1.00 | | | | | |
| 2 Subject Line Sentiment | .16** | 1.00 | | | | |
| 3 Email Content Complexity | .01** | .00 | 1.00 | | | |
| 4 Subject Line Complexity | .02** | -.09** | .02** | 1.00 | | |
| 5 Email Content Emotionality | .12** | .04** | .50** | -.08** | 1.00 | |
| 6 Subject Line Emotionality | .01* | .08** | .02** | .08** | .10** | 1.00 |

*p<.05; **p<.01.

Table 4. Correlating semantic variables of emails bodies and subject lines.

As the table shows, correlations are positive and rather significant at the author level, meaning that the social actors are consistent in their average sentiment, emotionality and complexity, both in the email bodies and in the subject lines. On the other hand, correlations at the email level are rather low and their significance is mostly driven by the large number of observations. It seems, in this case, that language is used differently in email bodies and subject lines. This finding does not exclude the possibility of subject lines to show predictive power in several social phenomena (e.g., Gloor, Fronzetti Colladon, Giacomelli, et al., 2017).

## 5. Discussion and Conclusions

The analysis of email communication of employees, as well as the study of interaction dynamics on social media, can offer unprecedented insights to generate business value (Elshendy, Fronzetti Colladon, Battistoni, & Gloor, 2018; Gloor, 2017).

In this paper, the authors analyzed the effects of node removal strategies, mostly intended as possible data cleaning strategies, on two real networks, one extracted from Twitter and the other from three years of email communications. They tested the stability of several network metrics – either widely used or revealing expressive power in previous research – both at the ego-network and at the whole-network level. The results show that the overall graph structure, as well as the centrality (degree and betweenness centrality and betweenness oscillations) and interaction metrics (average response times, activity, contribution index and nudges), tend to remain relatively stable, when removing spammers and extremely peripheral nodes. Only closeness centrality seems to have a general lower stability. These results are consistent with those of recent experiments of Fronzetti Colladon and Vagaggini (2017) – who analyzed intranet social networks –, but partially refute the expectation that removing spammers from the analysis alters the network structure. A potential explanation for this finding might be that spammers are a special set of social actors, who are not integrated in the interaction dynamics as much as their peers, because their messages are mostly unsolicited and often remain unanswered. The content of spam messages does not add meaning to the general discourse and is therefore often ignored, with little impact on the network structure. Even when spammers have a large number of links – because sending emails to all employees, or mentioning many users in a tweet – these links are rarely reciprocated. In Twitter networks, spam bots might also remain isolated, as their messages are rarely answered, mentioned or retweeted. There might be exceptions to this argument, for example in cases where fake news have significant propagation (Amarasingam, 2011; Shao, Ciampaglia, Varol, Flammini, & Menczer, 2017) – an event that did not happen in the authors' dataset. Accordingly, future research is needed to address the effect of misinformation, intended as a specific category of spam on social media.

Even if outside the focus of this research, it is important to remember that, although spammers do not significantly alter the network structure, their presence can still generate organizational costs – for example they can cause interruptions or disturbance to workflows (Jackson et al., 2001; Mark et al., 2008).

In this study, spammers were identified analyzing real world data from Twitter and email communications, as those social actors sending irrelevant or unsolicited messages – overcoming one of the limitations of simulated approaches, where it can be more difficult to differentiate between spammers and overly connected nodes. The authors found that one of the best possible strategies for data cleaning and graph simplification is to remove the identified spammers and the least connected nodes. Since communication networks rely on connectivity for their functioning, if vertices are removed, the length of the average shortest paths will increase – with possibilities of disconnections and interruption of communications among peers (Newman, 2010). This can be a major problem when targeting generic hubs, but is much less relevant when eliminating overly connected spammers: if eliminating a spammer would produce some isolates, the information lost would not be relevant. This way, one can reduce the computation time of network metrics, as well as reduce the distortion of insights coming from the network analysis. Such a strategy minimally affects the network metrics, without big variations in overall structure and individual scores of the remaining nodes. By contrast, focusing on the top connected nodes proved to be a more altering strategy, which may significantly change the social network structure and consequently the connection patterns between social actors. This is consistent with previous research addressing the problem of hub removal in scale-free networks (Albert et al., 2000; Callaway et al., 2000). In other words, simplifying the graph before carrying out a network analysis seems to be reasonable; when removing spammers and loosely connected nodes the social network structure, response dynamics and content seem to remain relatively stable (except closeness centrality and average degree). On the other hand, the same results support ignoring spammers without filtering them out, since the authors found that the impact of these social actors on network metrics is often negligible.

Spammers can be identified by many different strategies, depending on the context available, and the goals of the analysis (Bhowmick & Hazarika, 2018; Xu et al., 2016). The authors' findings suggest that contribution index might be a valuable metric for filtering out spammers, for instance when the content of messages is unknown: in the case study networks, nearly all spammers had a contribution index above 0.8.

A further contribution of this study was to consider semantic variables (language sentiment, emotionality and complexity) together with social network variables which also were stable except when removing the top nodes from the email network. In several real-world cases, privacy issues can limit data collection possibilities, as many companies deny fetching email bodies while studying their internal communication networks. For this reason, the authors tested the correlation of the semantic variables when calculated at the email and at the actor lever. They found that sentiment, emotionality and complexity are weakly correlated at the email level, but are strongly correlated at the actor level; meaning that the characteristics of the language used by a person are, on average, similarly reflected in his/her subject lines and email bodies. On the other hand, when considering single emails, one should treat subject lines and email bodies separately.

Generally speaking, every analyst who wants to carry out a graph simplification by means of node removal, should pre-emptively study the effects of such intervention both on the overall network structure and on the node-level variables. This paper provides examples and methods to test such effects. The main contribution is twofold. Firstly, this work extends the research about robustness and stability of social network metrics, when applying graph simplification strategies (Costenbader & Valente, 2003; Fronzetti Colladon & Vagaggini, 2017). Real networks are considered, overcoming some of the limitations of a simulated approach where for example the identification of real-world spammers is not possible. Contrary to studies mainly focused on centrality measures (e.g., Costenbader & Valente, 2003), this research also considers metrics of interaction over time (average response times and nudges) and the use of language of social actors. Secondly, this paper provides useful insights to guide those researchers (or business analysts) who study social media and email communication networks, in order to obtain new insights which can ultimately affect business value (Gloor, 2017). When using big data analytics, graph simplification and the removal of spammers can be necessary to reduce computational complexity and to clean the results of the analysis from distortions. This paper shows that removing spammers is relatively safe, as it does not lead to a significant change in social network metrics – both in email and Twitter networks. Moreover, this work presents a methodology, which can be replicated by network researchers when applying other removal strategies, to evaluate their impact and appropriateness.

Lastly, this paper suggests that a network analysis is still possible, even in those cases where the removal of spammers is unfeasible (as they are for example not identifiable): removing spammers produced a negligible change in relative actor scores – for all the social network metrics considered, except for closeness and average degree.

This research has some limitations originating from the choice of analyzing real world networks instead of using a simulated approach. While this choice allows a precise identification of the spammer nodes, which would not be so precise in a simulated experiment, it constrained the number of networks the authors were able to analyze. Moreover, both networks considered have a degree distribution which is approximately scale-free. Future research could explore other social network structures – such as random networks – considering a larger number of graphs and also testing different sets of network metrics developed for big data analytics (e.g., Chang, 2018). Additional ways of analyzing language could also be explored, such as novel ways to calculate sentiment and emotionality (e.g., Karyotis, Doctor, Iqbal, James, & Chang, 2018). In addition, it would be interesting to analyze interaction patterns extracted from other contexts or social platforms, such as face to face advice networks in the workplace, Facebook networks, or graphs extracted from online forums or knowledge sharing applications. Our methodology could be further tested on social clouds, where large amount of data can be stored or exchanged (e.g., Chang, 2017). In general, our approach can be replicated in many different contexts and with new metrics, to assess their robustness.

To conclude, while spammers are a nuisance, and indiscriminately sent "hurrah" messages of the CEO can lead to employee dissatisfaction (Kellaway, 2017), they at least do not seem to influence network characteristics. This is also true for indiscriminately sent tweets, which, through hitting friends and foes alike, will not change the overall Twitter network properties. This means that most metrics of social network structure, dynamics, and content are robust with regards to spammers.

**Acknowledgements**

The authors are grateful to Mattia Iovino for carrying out a preliminary pilot version of this study in his bachelor's thesis.

**Funding**

This research did not receive any specific grant from funding agencies in the public, commercial, or not-for-profit sectors.

*Complex Systems ECCS '06*. Oxford, UK.